# Atomic-scale mapping of superconductivity in the incoherent CDW mosaic phase of a transition metal dichalcogenide

Sandra Sajan[1,+], Haojie Guo[1,+], Tarushi Agarwal[2], Irián Sánchez-Ramírez[1], Chandan Patra[2], Maia G. Vergniory[1,3], Fernando de Juan[1,4], Ravi Prakash Singh[2] and Miguel M. Ugeda[1,4,5]*

[1]*Donostia International Physics Center, Paseo Manuel de Lardizábal 4, 20018 San Sebastián, Spain.*
[2]*Department of Physics, Indian Institute of Science Education and Research Bhopal, Bhopal 462066, India.*
[3]*Max Planck Institute for Chemical Physics of Solids, 01187 Dresden, Germany.*
[4]*Ikerbasque, Basque Foundation for Science, 48013 Bilbao, Spain.*
[5]*Centro de Física de Materiales, Paseo Manuel de Lardizábal 5, 20018 San Sebastián, Spain.*

*\* Corresponding author: mmugeda@dipc.org*

[+]*These authors contributed equally to this work.*

*The emergence of superconductivity in the octahedrally coordinated (1T) phase of $TaS_2$ is preceded by the intriguing loss of long-range order in the charge density wave (CDW). Such decoherence, attainable by different methods, results in the formation of nm-sized coherent CDW domains bound by a two-dimensional network of domain walls (DW) - mosaic phase -, which has been proposed as the spatial origin of the superconductivity. Here, we report the atomic-scale characterization of the superconducting state of 1T-TaSSe, a model 1T compound exhibiting the CDW mosaic phase. We use high-resolution scanning tunneling spectroscopy and Andreev spectroscopy to probe the microscopic nature of the superconducting state in unambiguous connection with the electronic structure of the mosaic phase. Spatially resolved conductance maps at the Fermi level at the onset of superconductivity reveal that the density of states is mostly localized on the CDW domains compared to the domain walls, which suggests their dominant role in the formation of superconductivity. This scenario is confirmed within the superconducting dome at 340mK, where superconductivity is fully developed, and the subtle spatial inhomogeneity of the superconducting gap remains unlinked to the domain wall network. Our results provide key new insights into the fundamental interplay between superconductivity and CDW in these relevant strongly correlated systems.*

A fascinating enigma in quantum matter is the intricate interplay between superconductivity and charge density waves, two collective states of apparent opposite nature concerning their dynamic and static electronic character, respectively. While the coexistence of these electronic phases is frequent, a complete understanding of the microscopic mechanisms governing their interplay remains often sorely lacking. In this arena, transition metal dichalcogenide (TMD) metals are ideal playgrounds for the exploration of this mutual interaction since their relatively simple atomic and electronic structure leads, however, to a rich variety of collective electronic orders. Remarkable examples here are the octahedrally coordinated (1T) TMD metals 1T-TaS$_2$ and 1T-TiSe$_2$, which show similar electronic phase diagrams. While none of these CDW materials are superconducting in their pristine form, superconductivity can be induced in both cases by similar methods such as electrostatic doping[1,2], pressure[3,4] and chemical substitution/intercalation[5–10]. These methods induce a commensurate (C) to incommensurate (IC) CDW transition that precedes the emergence of the superconductivity. The incommensuration leads to the formation of CCDW domains separated by sharp, interconnected DWs, which have been invoked as the driving force and spatial origin of the superconductivity[2–4]. Despite the efforts to elucidate the origin of superconductivity in these 1T-TMD metals[2,9,11–14], direct experimental evidence clarifying the specific impact of DWs in each system on the superconductivity is absent. Similarly, the microscopic nature and properties of the superconducting state in these materials remains barely explored.

In 1T-TaS$_2$, the emergence of superconductivity upon external actuation (doping, pressure, etc.) is particularly intriguing due to its initial insulating ground state at low temperatures, where CCDW order is developed with a ($\sqrt{13} \times \sqrt{13}$)-R13.9° superstructure (Fig. 1a). The origin of this insulating behavior has largely been attributed to a Mott state caused by strong electron correlations in the CDW[15–19]. More recently, however, this interpretation has been questioned, and the CDW interlayer coupling has been pointed as the origin of the insulating gap[19–25]. Nevertheless, the application of the above-mentioned methods leads commonly to the metallization of the material, which enables the development of superconductivity with optimal critical temperatures ($T_C$) ranging from 1-6 K according to recent transport measurements[1,3,7] (Fig. 1b). Despite the significant discovery of superconductivity in this material, its fundamental nature, including the symmetry of the order parameter, and the potential role of the DW network in its development, remains largely unexplored.

Here, we report the microscopic characterization of the superconducting state of 1T-TaSSe, a related compound of 1T-TaS$_2$ that hosts and optimizes superconductivity upon selenization. The gradual selenization[7,8] - isovalent substitution of S by Se - of 1T-TaS$_{2-x}$Se$_x$ leads to the melting of the CCDW for x > 0.8 in favor to the incommensurate CDW - also referred as the mosaic phase, followed by the emergence of superconductivity between 0.9 < x < 1.6 - see Fig. 1b. By means of scanning tunneling microscopy/spectroscopy (STM/STS) experiments, we first probe the electronic structure of the CDW mosaic phase that coexists with the superconducting state. Our conductance maps at the onset of the superconducting state (T = 2 K) show that the density of states (DOS) available for the Cooper pair condensation in the mosaic phase is rather concentrated at the coherent CDW domains, and, therefore, the presence of DWs is minor. We use high-resolution STS and point-contact spectroscopy at T = 0.34 K to show that superconductivity is held by a single gap of BCS character of $\Delta_{BCS}$ = 0.24 ± 0.02 meV. Spatial mapping of the superconducting gap and the coherence peaks indicates that the superconductivity is homogeneously extended regardless of the presence of DWs. The presence of domain walls, assuming them uncorrelated from layer to layer, leads to the randomization of the stacking of the CDW, which induces a metallic state through interlayer tunneling, thus enabling superconductivity. Our findings provide new perspectives on the interplay between the superconducting and CDW orders in this model 1T-TMD material, imposing stringent constraints on the origin of superconductivity.

We performed our experiments in high-quality 1T-TaSSe crystals exfoliated in ultra-high-vacuum (UHV) conditions prior to the STM/STS inspection (see methods). Figure 1c shows a typical topographic image of the uppermost 1T layer of TaSSe, where the CDW mosaic phase of 1T-TaSSe is evident. The size of the coherent CDW domains is variable but limited to a few tens of nanometers at most for this S:Se ratio. The network of domain walls (light 1D features in Fig. 1c arises from both rotational (27.8°) and translational disorder between the CDW domains, as shown in Figs. 1d and 1e, respectively. While the DW network is only visible in the uppermost layer in STM images, it is present in the layers underneath (Fig. 1f). This is expected to introduce a large vertical disorder in the CDW stacking sequence with significant impact on the electronic structure.

First, we characterize the large-scale electronic structure of the mosaic phase by d$I$/d$V$ spectroscopy. The orange d$I$/d$V$ spectrum shown in Fig. 2a qualitatively captures the essential features observed in this phase. While the occupied state region (negative bias) shows a featureless, increasing density of states (DOS), in the empty state region (positive bias) two pronounced, wide peaks appear for energies $V_b < 0.6$ V. These peaks were found roughly equidistant but varying within $E_F < V_b < 0.6$ V among different locations, as exemplified by the gray d$I$/d$V$ spectrum in Fig. 2a (different location). The variation of these empty-state resonances presumably arises from the spatially changing vertical stacking of the CDW, which has similarly been reported in 1T-TaS$_2$ (ref. 23). To confirm this scenario, we have performed density functional theory (DFT) calculations of bilayer 1T-TaSSe for different CDW stackings (AA and AB), which reproduce the main qualitative features of the experimental DOS as well as the peak-position dependence with stacking (see SI.2). Specifically, the first empty state peak corresponds to a broadened antibonding state of the flat band due to the local dimerization, while the second peak corresponds to the onset of the conduction bands above the flat band. This observation reflects the randomization of the CDW stacking in the out-of-plane direction, which we argue is the origin of the metallization of this layered material, including the CDW domains.

To better understand the orbital character of the electronic structure, we performed spatially resolved d$I$/d$V$ mapping of the mosaic phase within ± 1 eV. The upper panel in Fig. 2a shows four representative conductance maps acquired at different energies. These measurements reveal that the conductance is mostly localized around the central atom and six nearest neighbors of the Star of David (SoD) in the energy range -0.35 V $< V_b <$ 0.55 V (shaded region in the d$I$/d$V$ spectra of Fig.2a, as shown in the two central conductance maps. Beyond this energy window at both polarities, the intensity localizes at the rim of the SoD (see SI for details). This orbital texture in 1T-TaSSe is qualitatively similar to that recently observed in 1T-TaS$_2$ (refs. 9,25), and denotes the predominant role of the $d_{z^2}$ Ta-derived orbitals around $E_F$ where superconductivity develops.

We now focus on the spatial localization of the DOS around $E_F$ in the mosaic phase. Our STS measurements at both CDW domains and DWs show that the DOS around $E_F$ is finite (metallic character), but significantly lower compared to that at larger/lower energies. The metallic character of the entire surface at $E_F$ enables both regions to participate in the formation of the superconducting state. Despite the complex and varying electronic structure of the DWs observed in our STS data (SI.4), a common characteristic is the nearly equal, or even lower, DOS of the DW network around $E_F$ as compared to that at the CDW domains. To confirm this observation, we mapped the zero-bias ($E_F$) conductance ($\propto$ DOS) in several regions of the mosaic phase in 1T-TaSSe at $T$ = 2 K, right above the onset of the superconductivity. A representative dataset is shown in Figs. 2b,c, respectively. Consistent with the spatially resolved STS shown in Fig. 2a, the DOS at zero-bias in Fig. 2c is mostly located on SoD clusters (bright spots). The yellow lines approximately represent the CDW domains and, therefore, the DOS at CDW domains and DWs can be visualized. As seen, the DOS is unevenly distributed in the SoD lattice, and relatively depleted at the DWs. This and similar measurements with even higher spatial resolution (SI. 5) rule out the DW network as the only spatial origin of the superconducting state. In fact, the much larger DOS at the CDW domains as compared to that at the DW network suggests a much larger contribution of the CDW domains in the Cooper pairing formation.

To characterize the superconducting state of 1T-TaSSe, we first probe the quasiparticle spectrum around $E_F$ via high-resolution STS using non-superconducting tips at $T = 0.34$ K. Figure 3a shows a typical tunneling conductance spectrum in this energy region, which features a single, superconducting energy gap consistent with BCS superconductivity, with a magnitude of $\Delta_{BCS} = 0.24 \pm 0.02$ meV. We point out that the conductance often shows a dip at $E_F$ within $V_b \approx \pm 1$ mV (SI.6) that is unrelated to superconductivity, which has been misinterpreted before as the superconducting gap[26,27]. This dip, of unknown origin to date, is a feature usually observed in STS experiments in TMD metals[28–30].

To further confirm the superconducting nature and BCS structure of the full gap observed in 1T-TaSSe, we performed point-contact spectroscopy experiments by probing the in-gap conductance from tunneling to contact regimes approaching the metallic tip toward the surface. The emergence of in-gap conductance is caused by Andreev reflections, which proves the superconductivity and provides key information about the pairing symmetry of the order parameter. Figure 3b shows the evolution of Andreev states in a pristine region of 1T-TaSSe. As seen, the Andreev states gradually fill the superconducting gap as the tunneling resistance decreases, and lead to a conductance smoothly proportional to $V_b$ within $\Delta$ with no further features. This is the expected behavior for a s-wave superconductor-metal interface within the Blonder-Tinkham-Klapwijk (BTK) model[31], where the Andreev reflection probability ($\propto dI/dV$) steadily increases as the barrier strength diminishes and is directly proportional to $eV_b$ (see SI.7). In summary, both the tunneling and Andreev spectroscopy data on 1T-TaSSe are consistent with a conventional s-wave pairing symmetry. To complete the characterization of the superconducting state, we tracked the superconducting gap dependence with temperature and out-of-plane magnetic field. Figure 3b shows the gap evolution with temperature, which renders a critical temperature of $T_c = 1.5 \pm 0.1$ K. This value, along with the superconducting gap found ($\Delta_{BCS}$ = 0.24 ± 0.02 meV), yields the ratio $\frac{2\Delta}{k_B T_C}$ = 3.7 ± 0.5, which sets 1T-TaSSe as a weakly coupled BCS superconductor. The evolution of the superconducting gap with the magnetic field is shown in Fig.3b, and shows an upper critical field of $B_{c2}$ = 3.4 ± 0.2 T. These critical values and the corresponding errors have been obtained by averaging over different datasets.

Having characterized the superconducting state in 1T-TaSSe, we investigate the impact of the CDW mosaic phase on it. To this purpose, we have spatially mapped the fluctuations of the superconducting order parameter in multiple regions at 340mK, our experimental base temperature. Figure 4a shows a representative region where the CDW domains and DWs are clearly identified. In this region, the superconducting gap shows sizable spatial fluctuations as shown in Fig. 4b that, however, are homogeneously distributed, including the CDW domains, and unrelated to the presence of DWs. We obtain the same result by analyzing the local fluctuations of the coherence peaks, whose amplitude is related to the degree of the development of the superconductivity. Figures 4c,d show the amplitude of the negative and positive coherence peaks mapped over the same region in Fig. 4a, respectively. As in the case of the gap, both maps show homogeneously distributed spatial fluctuations where the presence of DWs is irrelevant. These local measurements confirm the homogeneous development of superconductivity on both CDW domains and the DW network, and suggest a negligible impact of the CDW incoherence on this phase.

In conclusion, our measurements provide a thorough microscopic characterization of the development of superconductivity in the family of octahedrally coordinated $TaS_2$, a structural polytype known to host intriguing phenomena arising from CDW-induced electron localization effects. Our results in 1T-TaSSe provide new insights into the nature of the superconductivity in the mosaic phase of $TaS_2$, and reveal that the key feature that enables superconductivity is the local randomization of the stacking of the CDW that induces a metallic state through interlayer tunneling. In contrast, the existence of domain walls within each layer has little influence in the superconducting state. These findings emphasize the three dimensional character of superconductivity and its robustness to the loss of long range CDW coherence, and impose stringent constraints in our understanding of superconductivity in $TaS_2$, in particular with respect to its interplay with the intriguing CDW mosaic phase.

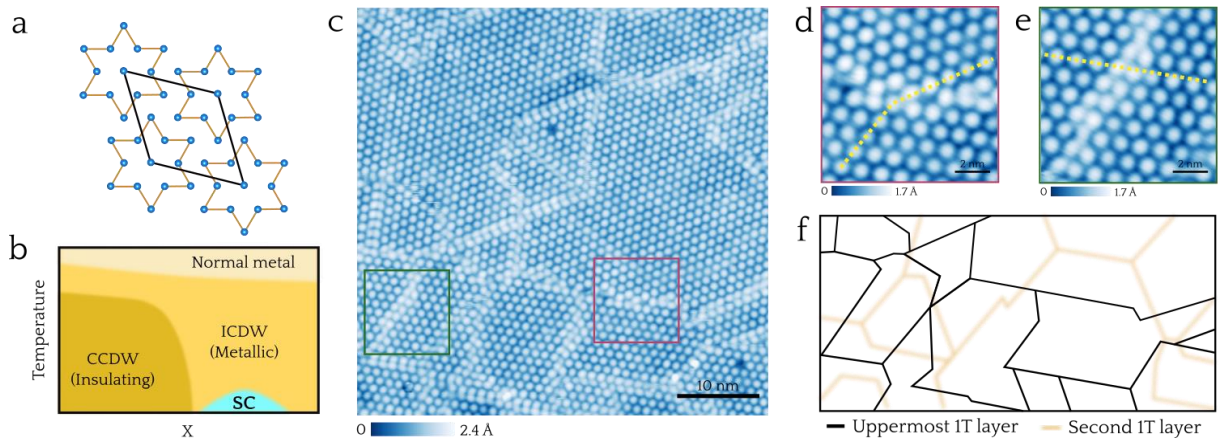

**Figure 1 | Mosaic CDW and superconductivity in 1T-TaSSe.** (a) Sketched unit cell (rhombus) of the CDW (√13 × √13)-R13.9º superstructure. Only Ta atoms are depicted. (b) Generic phase diagram of 1T-TMD metals where the superconductivity emerges followed by a CCDW-ICCDW transition. x variable in the abscissa axis may refer to pressure, doping, intercalation and, as in this case, Se content upon S substitution. (c) Large-scale STM image of the mosaic CDW phase in 1T-TaSSe ($V_s$ = 1 V, $I$ = 0.05 nA, $T$ = 0.34 K). (d) and (e), Zoomed regions from the original STM image in c showing two domain walls caused by rotational and translational mismatch, respectively. (f), Illustration of the vertical stacking disorder induced by the DW networks of each TaSSe layer. Only the first two layers are shown.

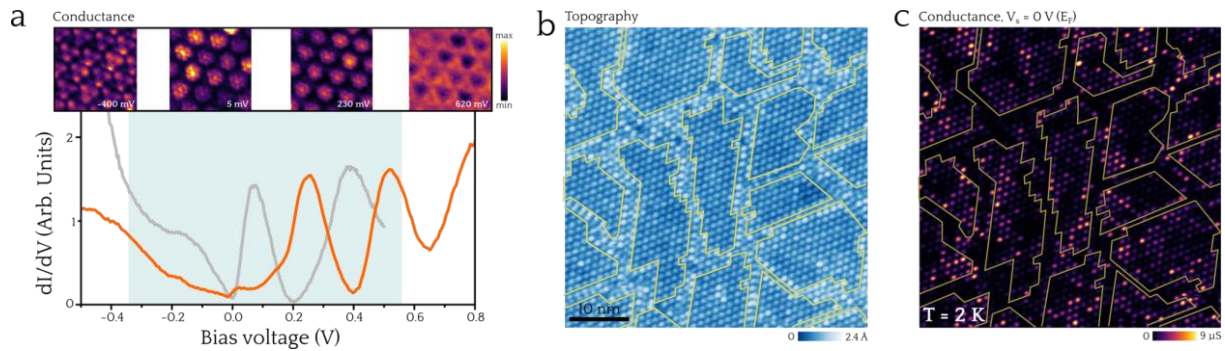

**Figure 2 | Electronic structure of the mosaic CDW phase.** (a) Upper panel, conductance maps taken in the same region of a CDW domain at different bias voltages. Lower panel, d$I$/d$V$ spectra acquired in different locations in CCDW domains of the mosaic phase ($V_{ac}$ = 5 mV, $T$ = 4.2 K). (b) Topography ($V_s$ = 0.37 V, $I$ = 0.3 nA) and (c), corresponding zero-bias conductance map of the mosaic phase acquired at $T$ = 2 K, slightly above the superconducting $T_C$ ($V_{ac}$ = 50 µV). The yellow contours denote approximately the boundaries between the CDW domains and DWs.

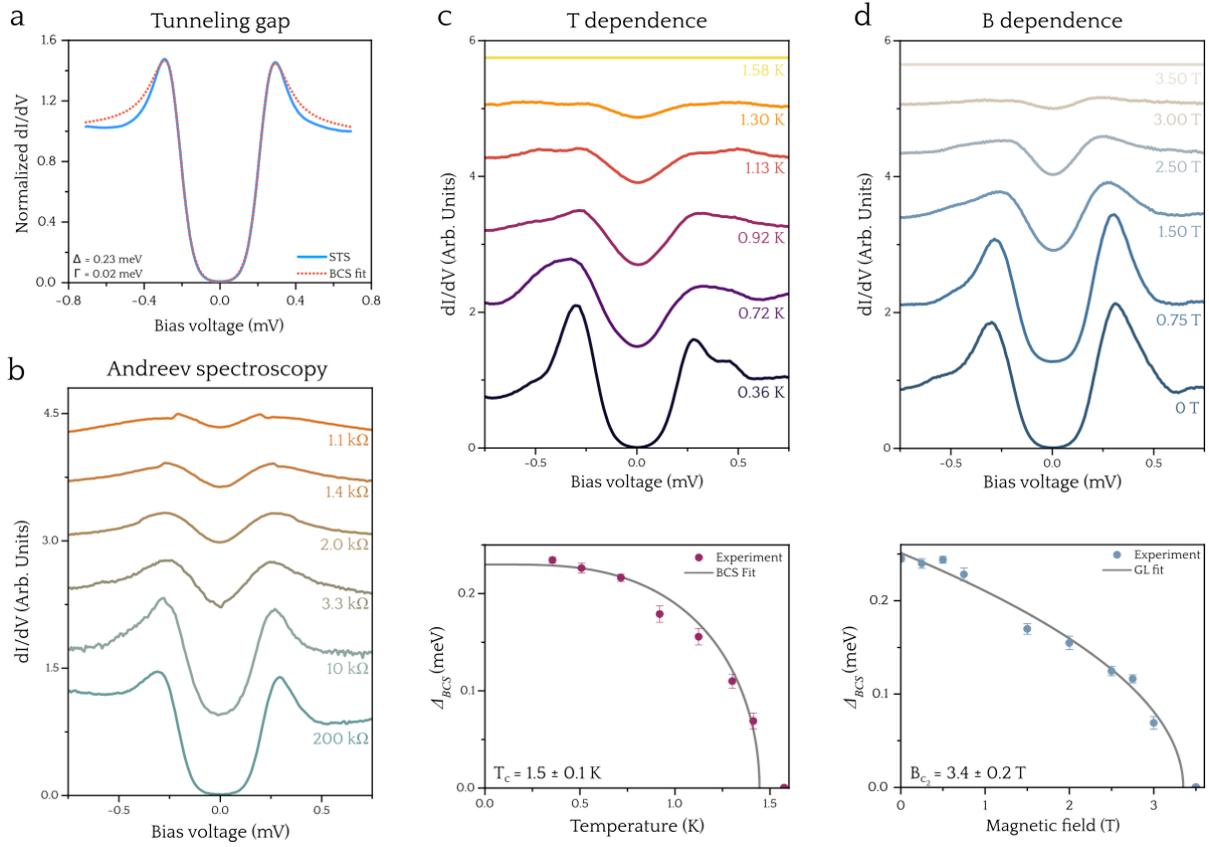

**Figure 3 | Nature of the superconducting state in the mosaic CDW phase.** (a) d$I$/d$V$ spectrum (spatially averaged over a region of 26 x 26 nm$^2$) showing the superconducting gap of 1T-TaSSe (blue curve). The dotted curve shows the Dynes function fitting employing an isotropic *s*-wave symmetry for the SC order parameter, yielding a SC gap size $\Delta = 0.23$ meV and a quasiparticle lifetime broadening $\Gamma = 20$ μeV. (b) Andreev spectroscopy spectra as function of the tunneling junction resistance ($V_{ac} = 14$ μV, $T = 0.34$ K). (c) Top panel: temperature evolution of the SC gap ($V_{ac} = 20$ μV). Bottom panel: BCS SC gap as a function of temperature. The solid gray line indicates the fitted line using an interpolated BCS formula. (d) Same as in (c) but for magnetic field dependence. The fitting here uses a phenomenological equation within the Ginzburg-Landau theory ($V_{ac} = 20$ μV, $T = 0.34$ K).

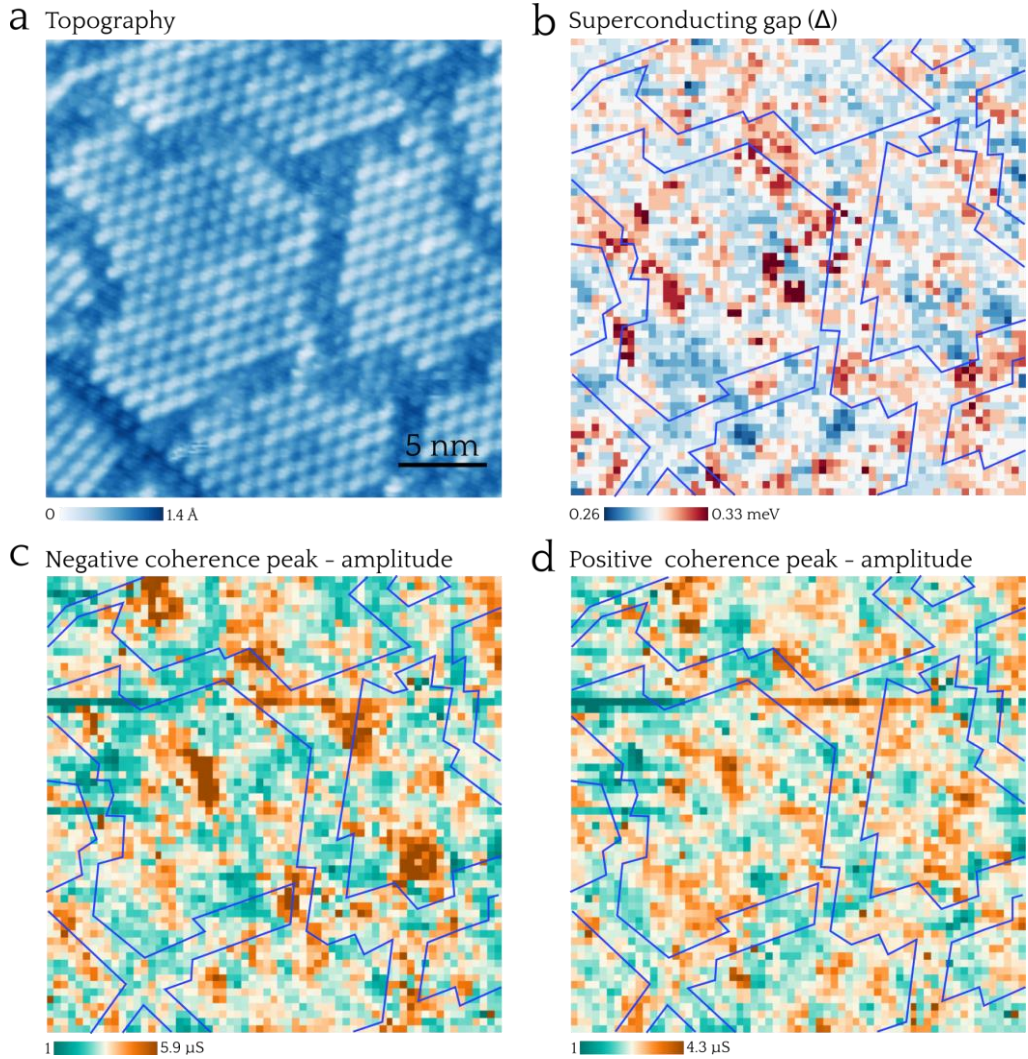

**Figure 4 | Spatial fluctuations of superconductivity in the mosaic CDW phase.** (a) STM image of the mosaic CDW phase in 1T-TaSSe where the superconducting fluctuations are probed ($V_s$ = -0.09 V, $I$ = 0.09 nA, $T$ = 0.34 K). Simultaneous measure of the $\Delta(r)$ map (measured as energy distance between the coherence peaks) in (b), and tunneling conductance at the coherence peak's energies in (c),(d), in the region shown in (a) ($V_{ac}$ = 15 µV, $T$ = 0.34 K). The blue lines in (b)-(d) indicate the contours of the CDW domains to highlight the negligible impact of the mosaic phase features in the superconductivity.

**Notes**

The authors declare no competing financial interest.

**Acknowledgements**

M.M.U. acknowledges support by the ERC Starting grant LINKSPM (Grant #758558) and by the grant PID2023-153277NB-I00 funded by the Spanish Ministry of Science, Innovation and Universities. H.G. acknowledges funding from the EU NextGenerationEU/PRTR-C17.I1, as well as by the IKUR Strategy under the collaboration agreement between Ikerbasque Foundation and DIPC on behalf of the Department of Education of the Basque Government. F. J. acknowledges support of the grant PID2021-128760NB-I00 funded by the Spanish Ministry of Science and Innovation. M.G.V. and I.S.R. thanks support from the Deutsche Forschungsgemeinschaft (DFG, German Research Foundation) GA3314/1-1 -FOR 5249 (QUAST) and the Spanish Ministerio de Ciencia e Innovacion grant PID2022-142008NB-I0. S.S. acknowledges enrollment in the doctorate program "*Physics of Nanostructures and Advanced Materials*" from the "*Advanced polymers and materials, physics, chemistry and technology*" department of the Universidad del País Vasco (UPV/EHU).

# Supplementary information for

# Atomic-scale mapping of superconductivity in the incoherent CDW mosaic phase of a transition metal dichalcogenide

Sandra Sajan[+], Haojie Guo[+], Tarushi Agarwal, Irián Sánchez-Ramírez, Chandan Patra, Maia G. Vergniory, Fernando de Juan, Ravi Prakash Singh and Miguel M. Ugeda*

*Corresponding author:* [mmugeda@dipc.org](mmugeda@dipc.org)

[+]*These authors contributed equally to this work.*

**This PDF includes:**

## SI.1: 1T-TaSSe crystal growth, initial characterization and STM/STS measurements

1T-TaSSe single crystals were grown directly using chemical vapor transport method using Iodine as a transport agent. High purity Ta, Se and S powders at a nominal composition were thoroughly ground and sealed in an evacuated quartz tube. This assembly was placed inside a two-zone furnace to maintain a precise temperature gradient. Different temperature gradients were tested and the effective gradient for obtaining large size crystals was found to be 1000 °C - 940 °C. Large crystals with some goldish appearance were harvested from the cold zone, measuring approximately 4×2×0.3 mm in dimensions (Fig. S1a).

The crystal structure of 1T-TaSSe was identified by the X-ray diffraction analysis with Cu-Kα radiation, performed at room temperature (Fig. S1). The Rietveld refinement of the powder pattern results in the trigonal crystal structure with P-3m1 (164) space group and the lattice parameters $a = b = 3.417 \pm 0.006$ Å and $c = 6.126 \pm 0.006$ Å. The crystal XRD pattern of 1T-TaSSe shows peaks along *[00l]* planes, indicating *c*-direction as a growth direction. The stoichiometric ratio of elements is determined by Energy-dispersive X-ray spectroscopy (EDS) measurement which is Ta:Se:S = 1:0.9:1.05.

Superconductivity in 1T-TaSSe was determined by temperature dependent magnetization measurement. Magnetization measurement was carried out in zero field cooled (ZFC) and field cooled (FC) modes at $H = 1$ mT. The observed diamagnetic signal below $3.25 \pm 0.05$ K indicates the onset of superconducting fluctuations in 1T-TaSSe.

STM/STS data were acquired in a commercial UHV-STM system equipped with perpendicular magnetic fields up to 11 T (Unisoku, USM-1300). The measurements were carried out at temperatures between 0.34 K and 4.2 K. To avoid tip artifacts in our STS measurements, the STM tips were calibrated using a Cu(111) surface as a reference. We also performed careful inspection of the DOS around $E_F$ to avoid the use of functionalized tips showing strong variations in the DOS. The typical lock-in a.c. modulation parameters for low- and large-bias STS were 20-50 µV and ~1-5 mV at $f = 833$ Hz, respectively. We used Pt/Ir tips for the STM/STS experiments. STM/STS data were analyzed and rendered using WSxM software[1].

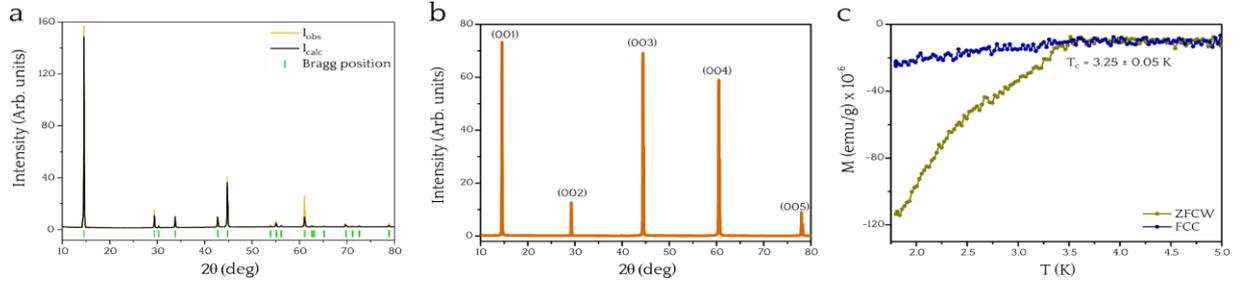

**Figure S1 | 1T-TaSSe crystal growth and x-ray diffractometry. a**, Crystal powder x-ray diffraction (XRD) pattern of 1T-TaSSe. **b**, The single crystal XRD pattern of 1T-TaSSe with peaks along (00l) directions. **c**, Magnetization measurement of the onset of the superconducting fluctuations.

**SI.2: DFT calculations of bilayer 1T-TaSSe**

All calculations were performed using Vienna *ab-initio* Simmulation Package (VASP)[2,3] v.6.2.1. with projector-augmented wave pseudopotentials within the Perdew Burke Ernzerhof parametrization[4]. Self-consistent calculations were found to be well converged with a 480 eV kinetic cutoff and a gamma-centered 15x15x1 k-mesh. T/T-TaSSe structure is built from 1T-TaSe$_2$ CDW √13x√13 layers from with 50% randomized S atoms in chalcogen positions[5]. Separation between layers is chosen to be commensurate with bulk and set to 6.29 Å for inter-layer Ta-Ta distances[6]. Projected band structures from Fig. S2 were obtained using PyProcar package in Python[7].

Figure S2 (a) and (b) display the A-atom-projected electronic band structure for the top 1T layer in both AA and AB stackings, respectively. The band structures for both configurations exhibit similar features: two almost flat bands, above and below the Fermi level, lying in the gap between two sets of dispersive entangled bands located at greater and lower energies.

The atom-projected DOS curves displayed in the upper and middle panels in Fig. S2 (c) show three sharp peaks. These peaks correspond, from left to right, to the two flat bands and the first set of entangled conduction bands. In order to facilitate the comparison with experimental d$I$/d$V$ results, the peaks associated with the upper flat band and the first group of conduction bands have been designated as $C_1$ and $C_2$, respectively, since they will be the only ones visible experimentally. The comparison between the DFT calculated DOS and the experimental d$I$/d$V$ measurements shows substantial agreement, in particular when examining the energy gap between the peaks labeled $C_1$ and $C_2$. The separation between the unoccupied flat band and the nearest conduction band in DFT is greater for AA stacking $\Delta E_{AA} \approx 0.25$ eV than for AB stacking $\Delta E_{AB} \approx 0.30$ eV; this trend is corroborated for d$I$/d$V$ measurements, where $\Delta E_{AA} < 0.30$ eV and $\Delta E_{AB} > 0.30$ eV. Even when these calculations reflect the main qualitative features of the surface DOS for a given stacking of the outermost layers, we caution that the finite DOS at the Fermi level is only recovered in a bulk calculation with a disordered stacking arrangement.

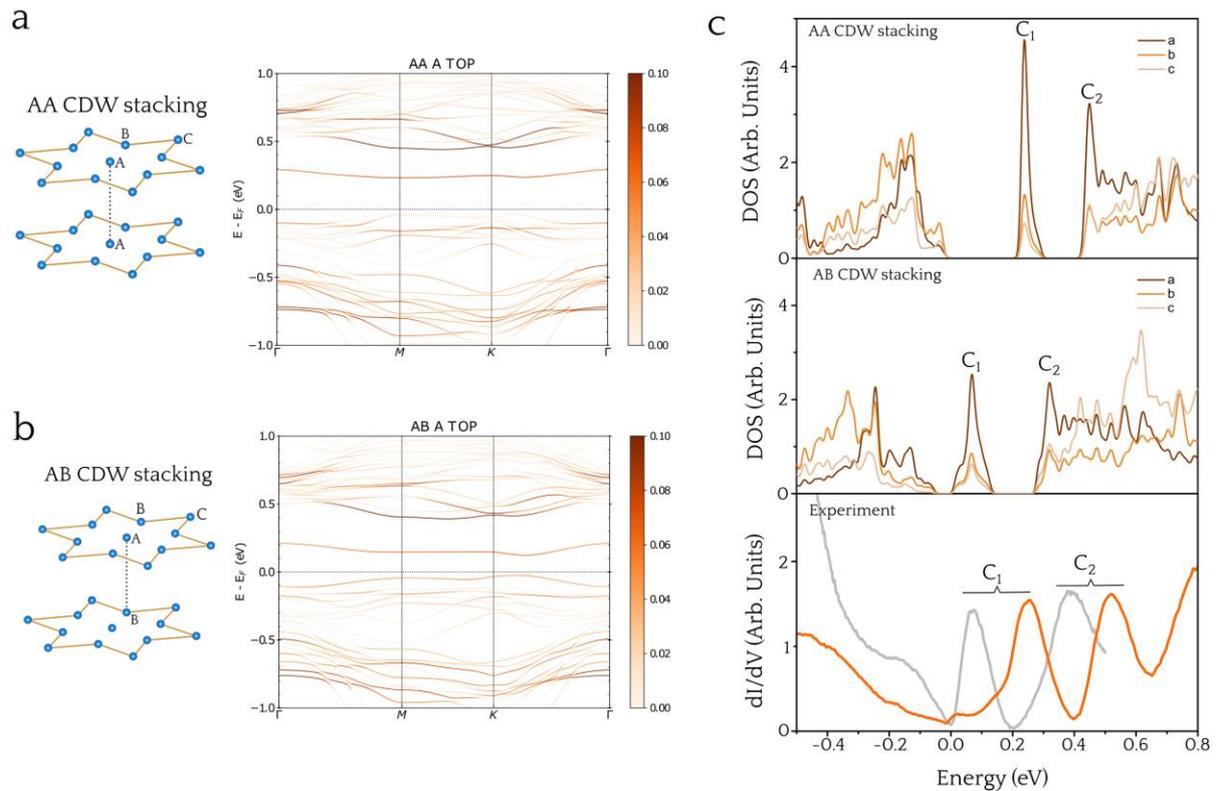

**Figure S2 | DFT calculations of bilayer 1T-TaSSe in the CDW state.** (a) and (b), Sketch (left) and calculated band structure of a bilayer of 1T-TaSSe in the CDW phase with AA and AB stacking, respectively. (c) Corresponding DOS of the calculated structures in (a) and (b) (upper and middle panel, respectively). The lower panel shows the d$I$/d$V$ spectra acquired in the mosaic phase and shown in Fig. 2 of the main manuscript.

## SI.3: Spatially resolved STS within CDW domains

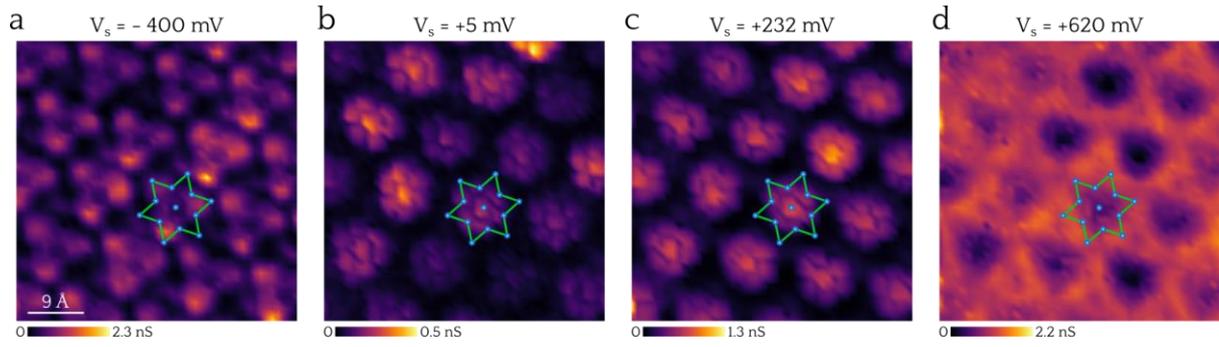

**Figure S3 | Spatial mapping of the orbital texture of the CDW of the 1T-TaSSe at selected bias voltage.** On each image (same as the one shown in Fig. 2a in the main text), we superimposed the SoD atomic model for the first chalcogen atoms layer. At conduction and valence band energy (a, d), the electronic distribution is dominated by triangles connecting the three chalcogen atoms vertices. Near the fermi level (b, c), the charge distribution is located mainly in the inner seven atoms, with the central one, that sits on top of the Ta atom below, being slightly brighter. Acquisition parameters for (a-d): $V_{ac} = 5$ mV, $T = 4.2$ K.

## SI.4: Tunneling spectroscopy on CDW domain walls

In general, the local electronic structure of the domain walls of the CDW mosaic phase is complex and strongly spatially inhomogeneous, as exemplified by the distinct d$I$/d$V$ spectra displayed in Fig. S4, acquired at the different domain walls. However, when compared to d$I$/d$V$ spectra simultaneously acquired on the CDW domains (blue curves in Fig. S4), two key observations emerge: the DOS is consistently similar at low bias voltage near the Fermi level, whereas significant differences manifest at higher energies in both conduction and valence bands. This suggests that the domain walls do not play a major role in the development of superconductivity in 1T-TaSSe.

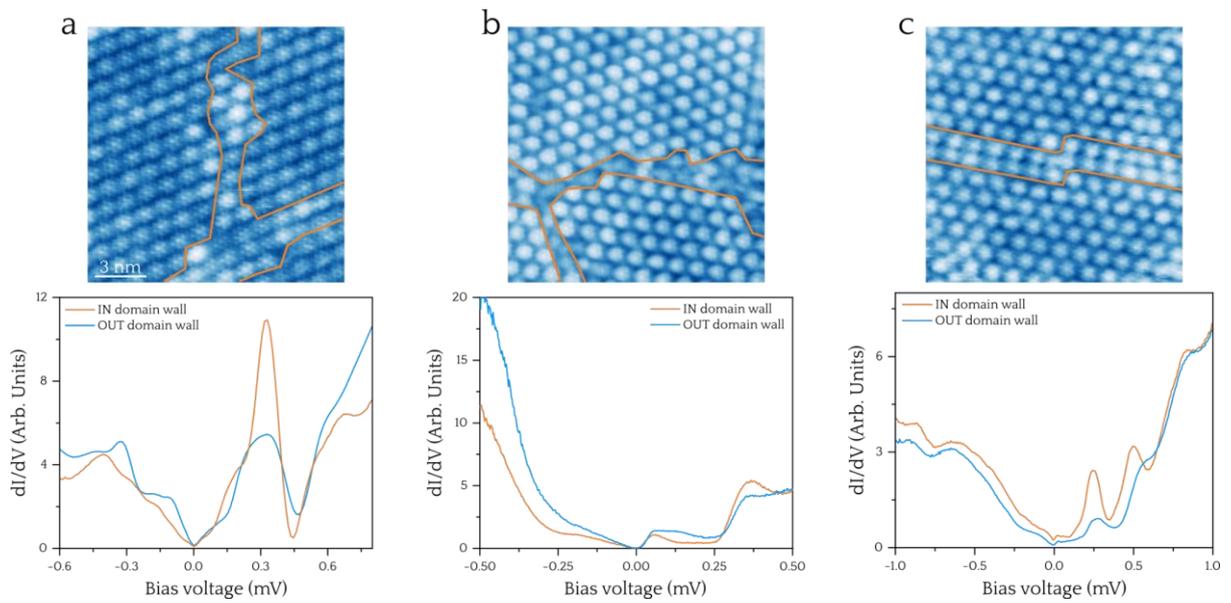

**Figure S4 | Large-scale electronic structure of CDW domain walls of the mosaic phase in 1T-TaSSe.** (a)-(c) Set of topography images of CDW domain walls along with selected d$I$/d$V$ curves acquired in (out) the DWs in orange (blue). Acquisition parameters: (a) $V_{ac}$ = 5 mV, $T$ = 2 K, (b) $V_{ac}$ = 3 mV, $T$ = 4.2 K and c, $V_{ac}$ = 5 mV, $T$ = 4.2 K.

**SI.5: High-resolution mapping of the DOS at domain walls above $T_C$**

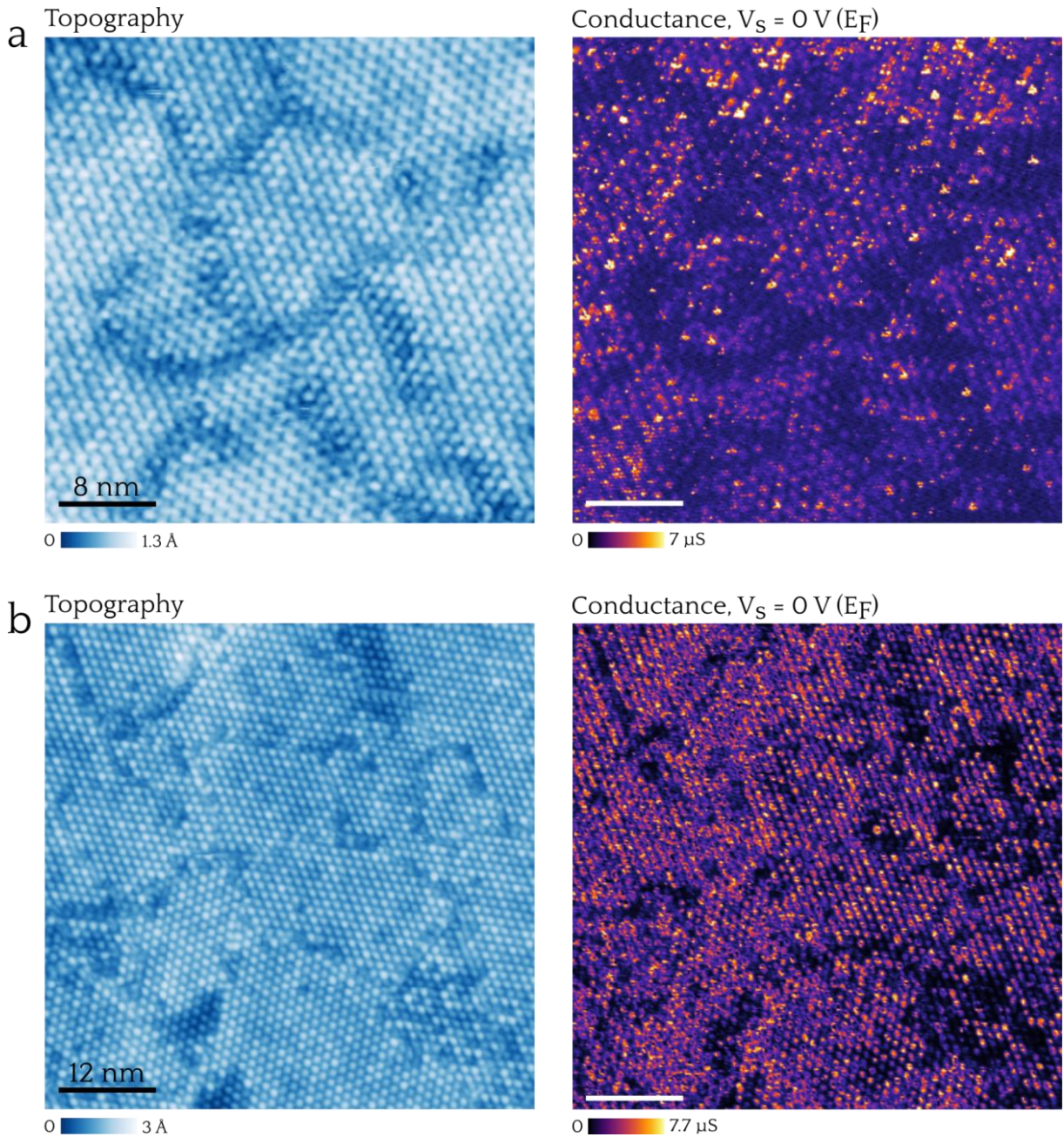

**Figure S5 | Density of states at the domain walls at the onset of superconductivity in 1T-TaSSe.** (a) and (b) show two different regions of the mosaic CDW phase in 1T-TaSSe where the topography (left images) and tunneling conductance at $E_F$ (V = 0 V) (right images) were measured at the onset of superconductivity ($T$ = 2 K). The depletion of DOS at the DWs at $E_F$ is evident. ($V_{ac}$ = 0.5 mV).

## SI.6: Low-lying electronic structure near the superconducting gap

In the quasiparticle tunneling spectra of the superconducting state of 1T-TaSSe, STS measurements seldom show an additional dip within the energy range of ± 0.75 mV to ± 1.5 mV. Three examples are shown in Figure S6a-c. This dip has been misinterpreted before as the superconducting gap[8,9]. If this dip was related to superconductivity, it would be consistently detected across the sample surface in STS measurements as it occurs with the inner gap at ± 0.23 mV. Instead, this is not the case and the dip is not systematically observed (Figure S6d-f), even when measured on the same area and with the same tip. Furthermore, the wide energy range in which we observed its onset rules it out as a superconducting gap. Lastly, nearly identical dip features have been observed in other non-superconducting TMD monolayers such as in single-layer $TaSe_2$ (NbSe$_2$) on BLG/SiC(0001)[10,11] and $NbSe_2$ on $WSe_2$[12]. In all cases, it has been demonstrated that the dip-feature is unrelated to superconductivity.

While the origin of this dip structure in TMD metals is not clear at the present time, we argue that it may likely be arising from disorder, tunneling effects or, even, a phonon-assisted inelastic channel feature. Further effort would be desirable to unveil the nature of this electronic feature common to most of the TMD metals.

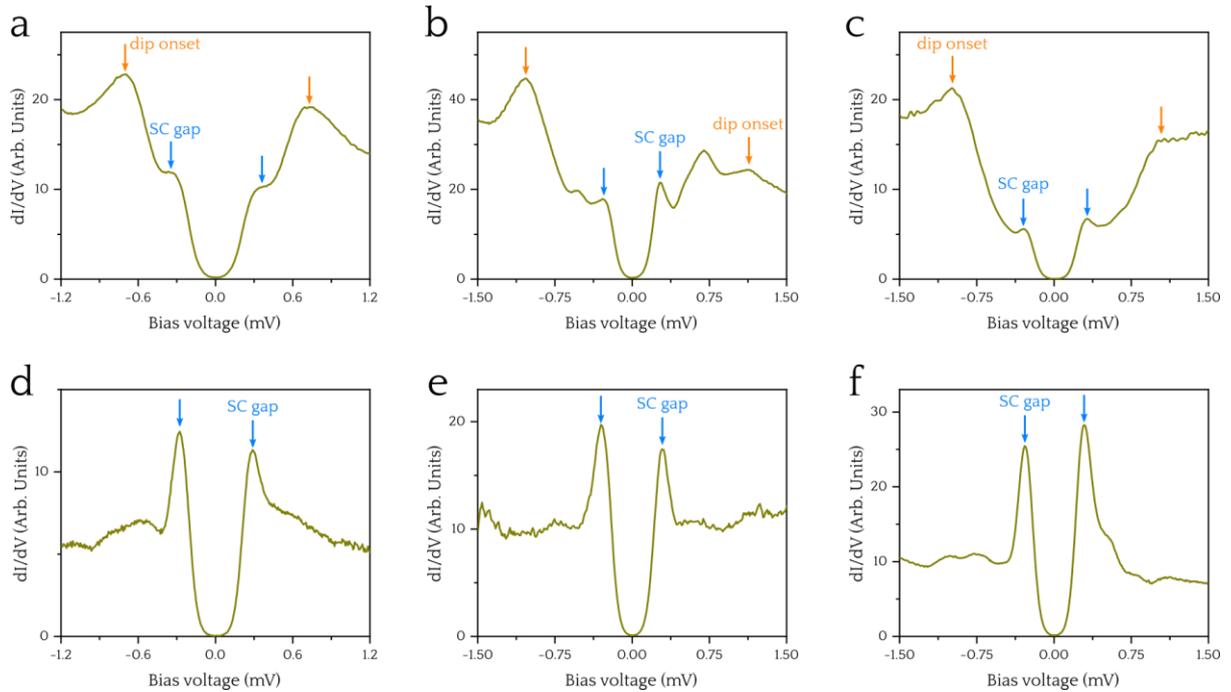

**Figure S6 | Dip feature in the low-energy electronic structure of 1T-TaSSe.** (a)-(c) d$I$/d$V$ spectra showing the additional dip at ± 1 mV (orange arrows), where the superconducting gap (blue) is embedded. (d)-(f) d$I$/d$V$ spectra showing the superconducting tunneling gap in the absence of the dip. ($V_{ac}$ = 20 μeV, $T$ = 0.34 K).

## SI.7: Blonder-Tinkham-Klapwijk model for s-wave superconductivity

Here we discuss the qualitative comparison between the Andreev spectroscopy data shown in Fig. 3b of the main manuscript (reproduced in Fig S7a) with the Blonder-Tinkham-Klapwijk (BTK) model using an isotropic superconducting order parameter (*s*-wave). Figure S7b shows a set of simulated Andreev reflection spectra for different tunneling barrier heights Z. As seen, the Andreev states gradually fill the SC gap homogeneously as the height of the barrier diminishes, which is consistent with our experimental observations. For Z = 0 (perfect transmission), the conductance develops a maximum flat conductance of $2G_N$, a value that we do not reach in our experiments likely due the finite barrier value even in the contact regime. It is important to notice that other symmetries of the superconducting order parameter give rise to more structured Andreev conductance shapes, which are distinguishable from that shown for an s-wave superconductor[13,14].

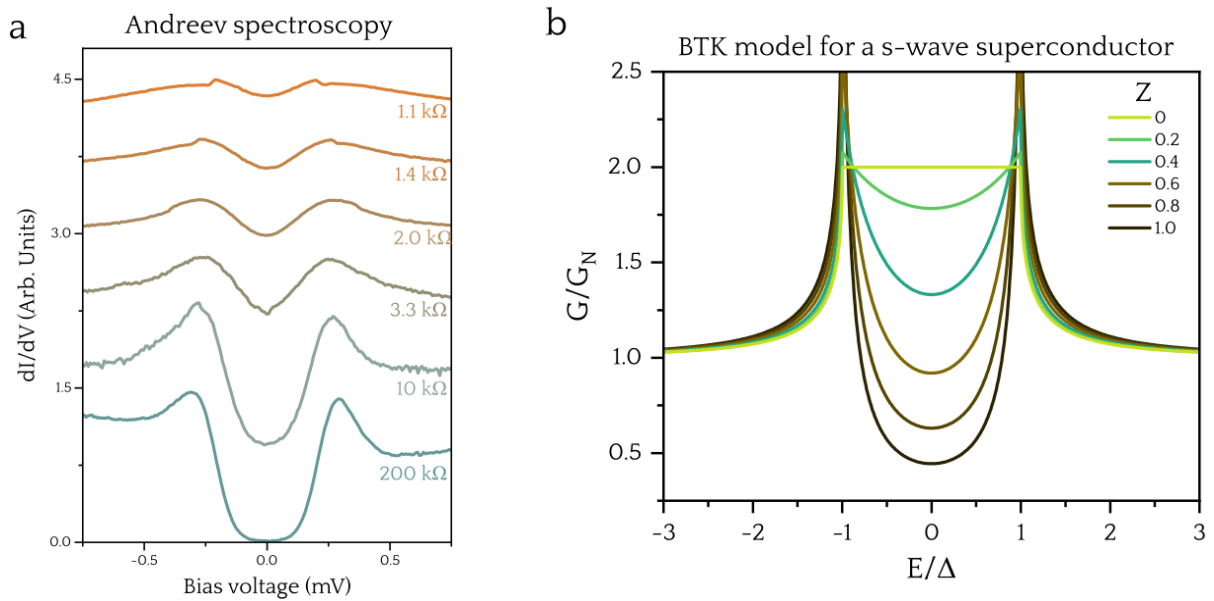

**Figure S7 | Andreev spectroscopy and the BTK model.** (a) Andreev spectroscopy spectra set shown in the main manuscript (Fig.3b). (b) Andreev conductivity simulated within the BTK model between a metal (non-superconducting) and a *s*-wave SC for various barrier heights Z.